# SkyServer Traffic Report – The First Five Years

Vik Singh, Jim Gray
Microsoft Research
Ani Thakar, Alexander S. Szalay, Jordan Raddick
The Johns Hopkins University
Bill Boroski, Svetlana Lebedeva, Brian Yanny
Fermilab

**Abstract** *The SkyServer is an Internet portal to the Sloan Digital Sky Survey Catalog Archive Server. From 2001 to 2006, there were a million visitors in 3 million sessions generating 170 million Web hits, 16 million ad-hoc SQL queries, and 62 million page views. The site currently averages 35 thousand visitors and 400 thousand sessions per month. The Web and SQL logs are public. We analyzed traffic and sessions by duration, usage pattern, data product, and client type (mortal or bot) over time. The analysis shows (1) the site's popularity, (2) the educational website that delivered nearly fifty thousand hours of interactive instruction, (3) the relative use of interactive, programmatic, and batch-local access, (4) the success of offering ad-hoc SQL, personal database, and batch job access to scientists as part of the data publication, (5) the continuing interest in "old" datasets, (6) the usage of SQL constructs, and (7) a novel approach of using the corpus of correct SQL queries to suggest similar but correct statements when a user presents an incorrect SQL statement.*

## 1 Introduction

### 1.1 Background

The multi-Terabyte Sloan Digital Sky Survey [1] – by far the largest digital astronomy archive to date [2] – is accessible online to astronomers and the general public via two Web portals. The *raw binary data* is available as flat files using wget/rsync from the Data Archive Server (DAS), and the distilled science parameters are extracted into the *catalog science archive* and available through advanced query interfaces from the Catalog Archive Server (CAS). The CAS is a collection of SQL Server databases [3] each storing a particular "release" of the SDSS data.

The study here analyzes CAS activity for the Early Data Release (EDR) and data releases 1 through 4 (DR1 – DR4). DR5 was just coming online as this study began. EDR was 80GB with 14M objects, 50K spectra. The later releases were 0.5TB, 1.0TB, 1.5TB, and 2.0TB. DR5 is 2.5TB with 215M photo objects, 0.9M spectra, and ~10B rows spread among ~400 tables [4]. DR8 is projected to be 2.9TB (see Figure 1.) The SkyServer offers HTTP, SOAP, SQL, and batch access to the CAS, and is really a federation of Websites that serve different communities and functions:

*SkyServer.sdss.org* or *cas.sdss.org*: a public Website offering access to the SDSS data, documentation on the data, and online-astronomy education in six languages (English, Japanese, German, Portuguese, Spanish, and Hungarian.)

*Collaboration and Astronomer portals*: separate Websites operated for members of the SDSS collaboration (restricted access) and other professional astronomers that allow longer-running queries on dedicated hardware. The user interface is streamlined for use by professional astronomers, and collaboration members usually have exclusive access to each data release for a few months prior to its public availability.

*CasJobs* (batch jobs for the (CAS): A public Web service that allows users to create a personal database (MyDB) on a server at Fermilab, upload personal datasets to it, and submit long-running programs and SQL queries that convolve MyDB data with the CAS datasets [5].

*Virtual Observatory*: A collection of Web services being developed by the Astronomy community as part of their efforts to build the World-Wide Telescope. It is not part of the SkyServer proper, but VO traffic appears in the Web logs.

## 2 SkyServer Hardware Infrastructure

The SkyServer is deployed on machines at Fermilab as described in Figure 1. The Virtual Observatory services are deployed on servers at The Johns Hopkins University (JHU). Since April 2001, we have been archiving the Web and SQL activity logs from the Fermilab and JHU servers. A collector running at JHU harvests the logs every few hours from across the Internet using a Web services interface offered by each SkyServer and CasJobs server (mirror servers in Europe, Asia, and South America have not been harvested so far). The harvested logs are aggregated in a publicly accessible database along with an activity summary [6]. Table 1 shows the overall statistics as of 1 July 2006, the corpus used here.

The logs have an opt-out privacy policy, but thus far no one has opted out [7]. Collaboration queries are hidden from public view but are included here because no one in the SDSS collaboration opted out of this study. Hence our database contains the full Web and SQL logs from Fermilab and JHU along with the analysis [8].

| **Table 1**: Overall statistics. | | |
|---|---|---|
|  | *Web* | *SQL* |
| *Log Start-Stop* | 2001/04/24 2006/07/01 | 2002/12/24 2006/07/01 |
| *Hits / queries* | 171,451,162 | 20,752,863 |
| *Page Views* | 62,481,516 | 16,123,600 |
| *Unique IP* | 925,666 | 19,497 |
| *Sessions* | 2,888,279 | 96,737 |

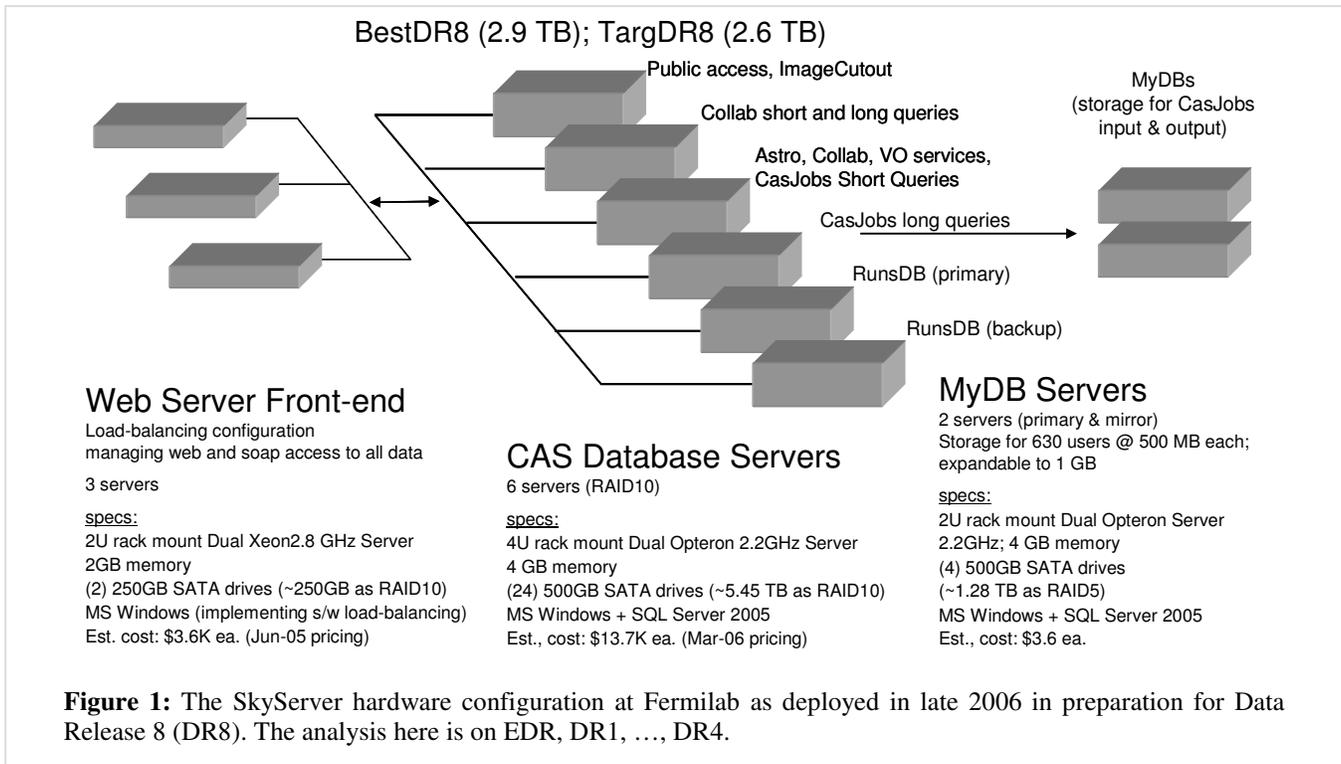

**Figure 1:** The SkyServer hardware configuration at Fermilab as deployed in late 2006 in preparation for Data Release 8 (DR8). The analysis here is on EDR, DR1, …, DR4.

### 2.1 Prior Work and Goals of This Study

Several prior studies used the public logs: R. Lees, using ThinSlicer™ built a datacube that allows easy analysis and visualization [9], R. Singh analyzed and visualized some session behavior [10, 11], and G. Abdulla analyzed term frequencies in the SQL logs [12]. In addition, T. Malik classified the structure of the SQL queries as part of her work on query-result caching [13]. This report analyzes long-term SkyServer usage patterns. Our goals are to:

(1) Characterize traffic volume and trends based on request-type (Web, Web-service, downloads, analysis…).
(2) Categorize the user population: astronomer, student, tourist, crawler, downloader, and others.
(3) Categorize the session behavior of each user segment.
(4) Characterize how users and bots use SQL.
(5) Assess the relative interest in datasets over time, in data within each data set, and perhaps make database design recommendations.

### 2.1 SkyServer Web and SQL Log Harvesting

The Web and SQL logs represent 75 system-years of activity collected from 60 server logs. They are a wonderful public resource, but they are not perfect. Each log has gaps. Some logs have records with incorrect or missing values due to bugs in our configuration or logging software. Much of the traffic is from crawlers and robot downloaders that swamp the traffic from mortals (people interacting directly with the Website.)

There are anomalies, like a Virtual Observatory registry manager that generated 42 million Web hits polling for changes to the registry. So, any analysis using the log data must be done with an understanding of the sites, and any results are approximate. We cleaned and normalized the HTTP and SQL logs and built ancillary data structures including:

*IP Name*: map the IP address to the institution owning that address block
*Sessions*: Organized time-sequences of requests from an IP address into sessions and computed statistics on each session
*Templates*: skeleton SQL statements with parameter numbers replaced with "#" and skeleton Web requests separating the *stem* (the url to the left of the "?") and *parameters* (the rest of the url)
*Agent Categorization*: for each web-agent string, we try to recognize the agent (e.g. MSIE or GoogleBot or Perl) and categorize it (e.g. browser or spider or bot).
*Page View flag***:** distinguish Web hits that are page views

The cleanup and normalization took several months effort. Figure 2 shows the resulting database design. The normalized database is 35GB (reduced from 180GB), accessible online [8].

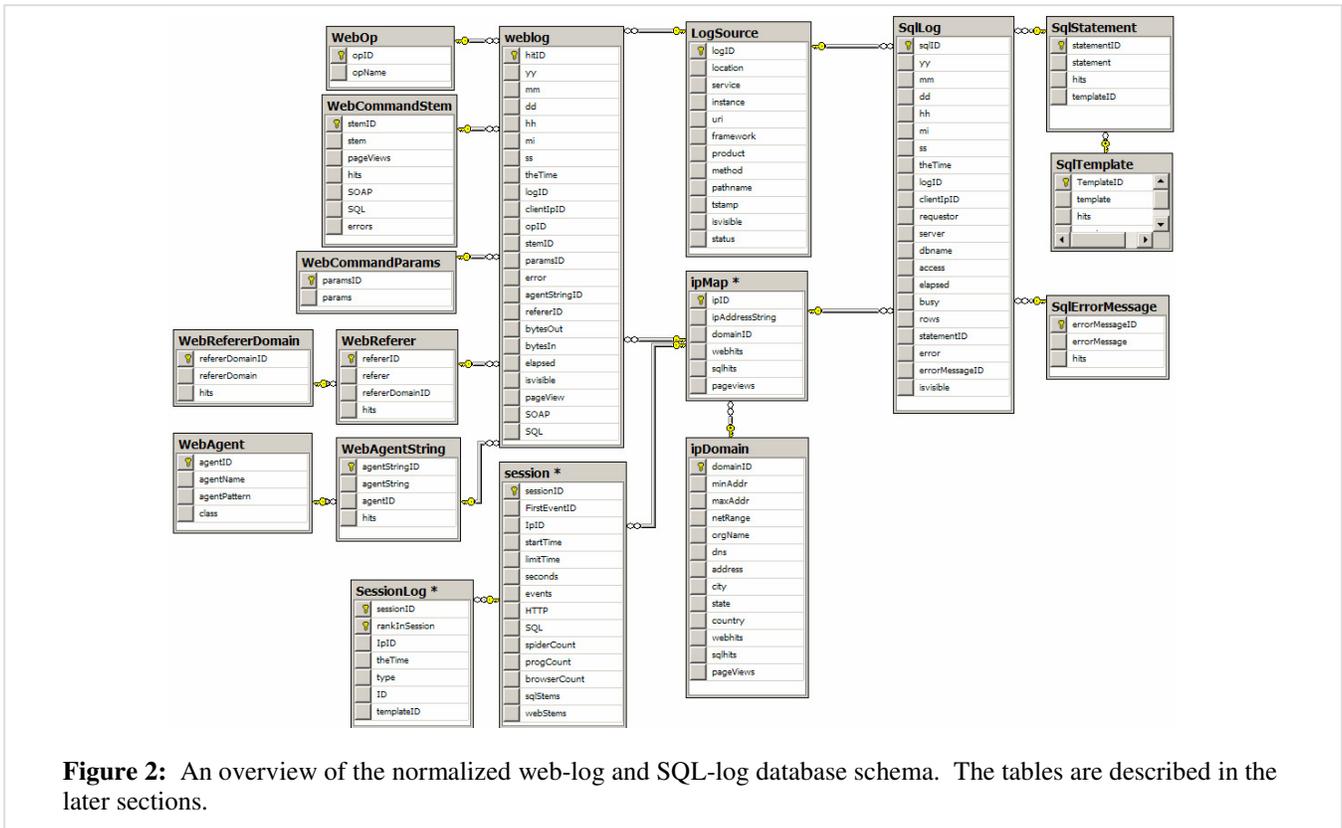

**Figure 2:** An overview of the normalized web-log and SQL-log database schema. The tables are described in the later sections.

## 3 Web-HTTP Traffic
### 3.1 Web Hits and Page View Traffic

Figure 3 summarizes the monthly Web traffic. The top line shows the total Web traffic on all servers measured in HTTP requests (hits). The Web-hit volume has doubled each year. The hits per month fit an exponential regression, (205% per year). In mid-2006, the logs averaged ~35K unique visitors and ~380K user sessions per month. As we will see, much of this growth is from programs (bots).

How many of these Web hits are just incidental to producing a Web page or Web-service answer? For example, displaying the SkyServer home page generates twenty hits if nothing is cached. The Web log has an entry for each request-reply pair (a hit), but many of those entries are either ancillary information (e.g., a .css style sheet for a Web page or a metadata .asmx file for a Web service), or are part of a larger package (e.g. one of the many .gif images on the home page), or are errors, or are redirects.

*Page views* measure how many answers the servers delivered to users or bots. We define a page view as any Web hit that (1) responds to a GET, HEAD, PUT, POST HTTP request or a SOAP request, (2) is not an error or redirect, (3) delivers information (status 200-299), (4) is not a noise type (e.g., .gif, .png, .txt, .css, .ico,..…), and (5) is not an administrative task from the BigBrother monitoring service or from the VO Registry Administrator.

Starting with 171M hits, 90% are the right request type, 4% of those return error, and 6% of the residue are redirects. Ignoring *BigBrother* and the *VORegistry* probes leaves 65 million page views. Figure 3 plots the page views, which display the same yearly doubling as Web hits.

There are daily, weekly and seasonal patterns: a mid-day peak, a Tuesday peak falling to a valley on the weekend, and relatively heavier traffic from November to March. Figure 3 shows the dominant patterns (1) year-over-year traffic doubling and (2) high short-term variability, with huge peaks and some lulls.

The statistics for http hits are 65% GET, 25% PUT, and 10% HEAD. Only 12% of the hits have a reference string saying where the request originated; of these, 98% of the referrals

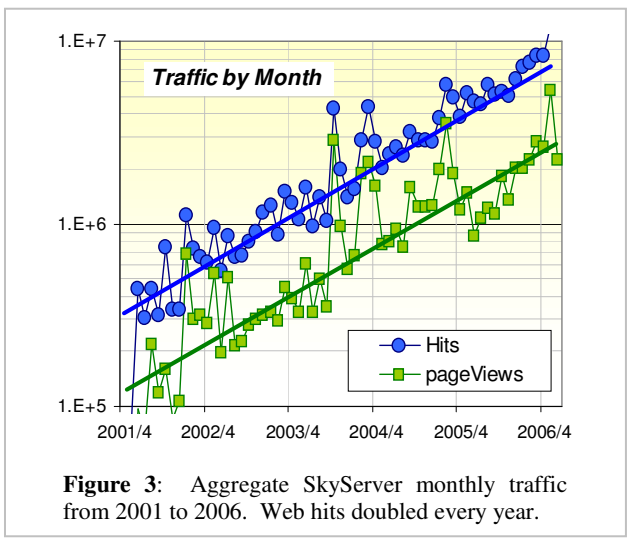

**Figure 3:** Aggregate SkyServer monthly traffic from 2001 to 2006. Web hits doubled every year.

are from SDSS sites, 1% are from Google (235k), and the remaining 1% are from 3,000 other sites.

Table 2 gives the relative frequency of the most popular Web page types – there were 78K hacker requests to execute various programs, many downloads of documentation, but most requests were for Web pages (asp, aspx) and Web services (asmx).

## 3.2 Session and User Segmentation

### 3.2.1 Clients

One of the main goals of this analysis is to characterize the way people and programs use the site. We segment human users into four broad categories:

*Scientists*: People using the site to analyze the SDSS data.
*Students*: People using the site to learn astronomy or other science topics.
*Tourists*: Users visiting the site out of curiosity.
*Administrators*: People, like us, analyzing site traffic.

We segment program behaviors as:

*Analyzers:* Programs running complex queries on SDSS data (e.g. CasJobs).
*Copiers:* Programs that systematically download parts of the SDSS database.
*Spiders:* Programs that crawl the Web pages to build an index.
*Administrators:* Programs that check site status, harvest Web logs, or maintain a registry.

We searched for ways to categorize page views into one of these eight categories; but had only modest success.

### 3.2.2 Categorizing Clients with Agent Strings

Users are anonymous. Each Web request carries an agent string that is supposed to tell what kind of agent browser or program generated the request. Sometimes the agent string tells who the client is (e.g. *Google*, *BigBrother*, *Perl*, *Safari*, *Firefox*); but agents often masquerade as Internet Explorer (*MSIE*) or some other popular browser in order to get certain behavior or in hopes of bypassing firewalls. So we are forced to classify users based on a combination of their (1) agent string, (2) IP address, and (3) behavior during a session. The one good thing is that a user's ipAddress is (by definition) constant during a session. However, a session may run several different programs and may include browser interactions; so, a session may have diverse agents, . In addition, the user's IP address may change from day to day. So, even these three attributes are only suggestive of the category that best describes a user or session.

Using the agent string classifies *some* of the hits as analysis clients (24 million), bot or spider clients (19 million), and administrative hits (18 million for *BigBrother*) with a residue of 118 million agent strings that look like browsers. We set the 42 million *VO-registry* probes to have a correct agent-string (VO-Registry rather than MSIE) leaving 76M hits. This classification, based purely on parsing the agent

| Table 2. Web-hit type frequency. | | |
|---|---:|---:|
| suffix | hits | Page views |
| asp | 64,128,683 | 60,111,219 |
| asmx | 43,728,961 | 1,680,388 |
| jpg | 22,794,275 | 0 |
| gif | 16,976,147 | 0 |
| aspx | 14,559,672 | 14,295,453 |
| htm | 8,777,611 | 5,144,895 |
| css | 3,255,012 | 3,379 |
| js | 1,527,566 | 0 |
| ICO | 1,446,242 | 0 |
| swf | 445,284 | 0 |
| txt | 411,916 | 0 |

string, is in the `WebAgent` table (Figure 2). It sub-classifies the bots into 78 groups (e.g. *Google, Slurp…*), programs into 10 groups (e.g. *python, java,..*) and the browsers into 11 groups (e.g. *Firefox, MSIE, Safari,..).* This parsing was helped by consulting IP registries [14]. Ignoring the administration traffic, the top two sub-groups are MSIE with 47 million hits and 19M page views and Python with 10 million hits and 9M page views.

### 3.2.3 Sessions

The logs record each client's *session* – the page view and SQL request sequence from an IP address. We arbitrarily start a new session when the previous page view from that IP address is more than 30 minutes old, i.e., a *think-time* larger than 30 minutes starts a new session. The thirty minute (1,800 second) think time is based on Figure 4 which plots page-view inter-arrival time frequency bucketed by powers of two. Thirty minutes captures 98% of them. The graph approximates a power law for times between 10 seconds and 10M seconds (100 days). Wong and Singh [11] chose the same 30 minute cutoff and we are told that MSN and Google use a similar heuristic.

As explained before, page views from *BigBrother* (17M views and 4.2M SQL queries) and the *VORegistry* administrator (42M views) are excluded. They comprise 34% of all hits and 21% of SQL queries, but they are just periodic probes of the Website, and they have traffic patterns we already understand. So they were excluded from sessions.

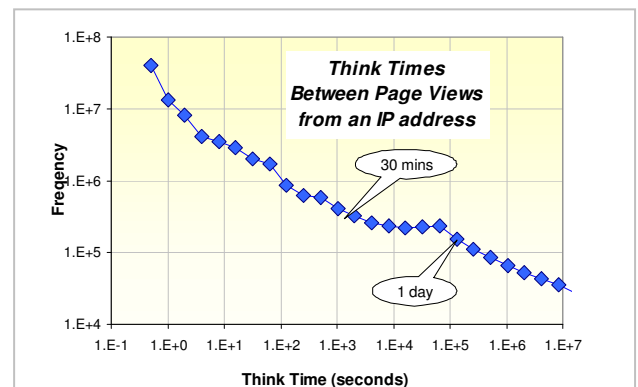

**Figure 4**: Think time (page-view inter-arrival time) from individual IP addresses bucketed in powers of 2 vs frequency. Most are short but some are more than a day. We arbitrarily chose 30 minutes as the session cutoff time.

This leaves 65,435,696 page views and 16,123,600 SQL queries in 2,985,016 sessions described by the `Session` table, and a `SessionLog` table indexed by `sessionID`, and `rankInSession`. The 65,435,696 `SessionLog` rows describe the session request sequences (pointers into `WebLog` and `SqlLog` tables) along with their timestamps, templates, and some summary information (Figure 2.)

### 3.2.4 Session Classification and Diversity

Our first task is to recognize and exclude spiders so that we can focus on the behavior of analysis, copy (data download), and human clients. If a client's *AgentString* declares a client IP address to be a spider or if the client IP address visits robots.txt then we declare all sessions from that IP address to be spiders. This eliminates 1.4M sessions, 14M page views and 328K SQL requests. Spiders were ½ the sessions, 18% of the page views, and 2% of the SQL traffic.

Recognizing the other categories is more difficult. We conjectured that people had irregular think times while programs would have a regular think-time pattern. Both those conjectures turned out to be false. Both people and programs seem to follow a power-law distribution of think times – so think-time is not a good way to distinguish them (see Figure 4.)

Figure 5 shows the frequency of session durations and session size (number of requests). Both graphs bucket the populations in powers of two (e.g. $\lfloor log_2(requests) \rfloor$ and $\lfloor log_2(duration) \rfloor$). The graphs show interesting patterns: Session lifetimes beyond a 1000 sec seem to follow an approximate power law behavior with a slope of -1.4. There is also a sharp cusp at short sessions. At the same time the number of requests per session follows a simple power law all the way – though SQL sessions tend to be longer than http-intensive sessions.

We conjectured that spiders crawl the Website and rarely re-visit the same page in a session. In line with this, we conjecture Copiers and Analyzers systematically crawl the database presenting the same request with different parameters, and we conjectured that people are a mix of the two behaviors; they visit several pages, may return to a page, and may dwell on a page as they submit queries.

These conjectures appear to be true in general. For example, consider sessions of Figure 5 that span more than 3 days (the ones lasting more than 250k seconds). Statistics for the top 5 are shown in Table 3. They came from five institutions doing systematic data downloads. Four of the institutions used the free-form SQL requests (`x_sql.asp` or `SkyQa.asp`) and two used the pre-canned SQL (`x_rect`) commands that do not record their SQL commands in the log. One uses the very popular GetJpegObj.asp that issues over a dozen different SQL calls to build an annotated JPEG image from the database, but that is just one Web command stem (virtually every SkyServer request has or more backend SQL actions in addition to generating a SQL Web log record). These sessions routinely had very few Web command stems (often one stem) and very few SQL templates. For example, the session with the most requests used x_sql.asp to ask the following question with 2.4M different number pairs, counting objects in each htm-range (spatial bucket):

```
select count(*)
from photoprimary
where (htmID >= # and htmID <= #)
```

| Table 3: Examples of 5 extremely long sessions | | | | |
|---|---|---|---|---|
| Hours | Pages | Web Cmd Stems | Free Form SQL Stmts | Methods (asp) |
| 140 | 2,479,279 | 1 | 3,572 | x_sql |
| 103 | 1,888,131 | 1 | 6,467 | x_sql |
| 368 | 1,448,323 | 2 | 1,098 | GetJpeg |
| 78 | 1,217,736 | 1 | 1 | x_rect |
| 100 | 1,171,158 | 1 | 2,571 | x_sql |

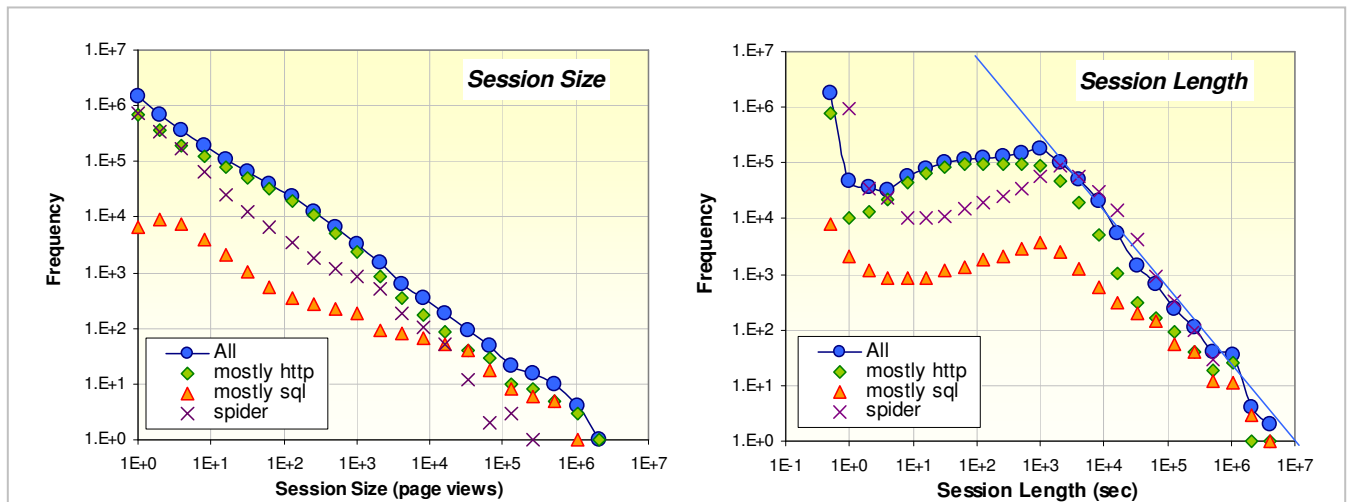

**Figure 5:** Session sizes (left) as measured in page views or SQL requests bucketed by powers of 2 (left) follow an approximate power law although SQL sessions tend to be longer and very long spider sessions are rare. The session lengths (duration in seconds) seem to have three different behaviors: sessions less than 3-seconds are popular, sessions lasting 3 to 1,000 seconds seem to follow one power low with a slight rise; then past 1,000 seconds session lengths seem to follow a second power law.

So, it had one Web stem …/x_sql.asp? and one SQL template (the statatement above). Yes; there are better ways to ask and answer this question; but this way works.

We define a session's *request diversity* as the ratio of requests to request stems (Web URLs to the left of the "?") plus SQL templates (statements with the non-identifier numbers replaced by "#"). The Web stems are in the `WebCommandStem` table and the SQL templates are in the `SqlTemplate` table of Figure 2. We expect spider sessions to have high diversity, copier sessions to have low diversity and people to have sessions with intermediate diversity. This hypothesis works very well at the extremes, but we were unable to get a crisp classifier from this approach. We found no clear break between the diversity of people, programs or spiders – the data looks like a continuum.

In the end, we despaired of an automatic way to recognize human users and bots based on statistics. Some statistics show clear bot behavior, 100 hour sessions or 1M page views in a session (!) but, the typical spider session is short 10 page views in 100 seconds – indeed that is why ½ the sessions are spider sessions (see Figure 5).

The best we could do in classifying sessions as mortal, was to find all sessions that were not administrative, not obviously a spider, not obviously a bot, lasted between one minute and 8 hours, and involved at least 4 page views or SQL requests. There were about ½ million such sessions.

Figure 6 shows the page-view and SQL request traffic for mortal sessions when averaged over 3 month windows. Web request traffic grew at 75% per year, while mortal SQL traffic quickly grew to ~30K requests/month and stabilized there. In comparison, the overall traffic doubles every year (Fig 3), thus there is a relative increase bots and spider usage. An interesting feature of the underlying data is that it seems to show a yearly trend with a dip in the summer and fall, and an increase in winter and spring.

## 4 Traffic by Source
### 4.1 Traffic by IP address

Each Web log entry and most SQL log entries carry an IP address. A reverse lookup converts this to the name of the institution owning the IP address. Unfortunately, many of the IP addresses resolve to large IP-address blocks that are "sub-leased" to many organization; so, the IP lookup maps to a large ISP – for example over 1,000 of the blocks mapped to the Amsterdam RIPE network which does not disclose it sub-leases. Nonetheless, a combination of automatic lookup and then some manual-resolution mapped most of the million IP addresses to about 11k IP domains.

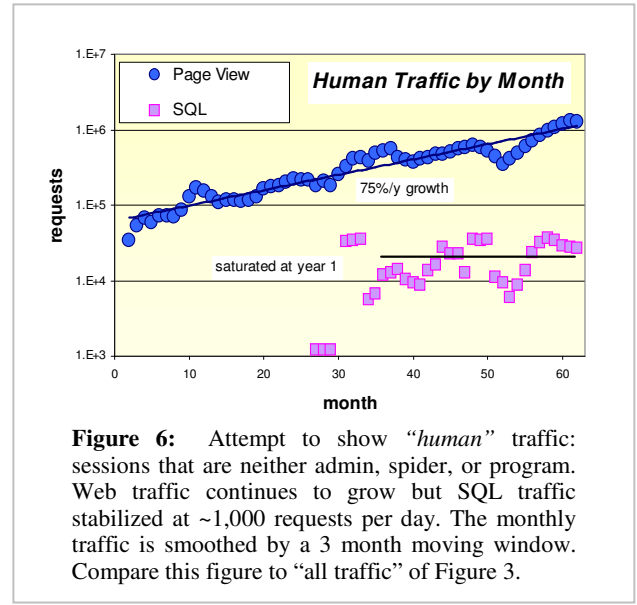

**Figure 6:** Attempt to show *"human"* traffic: sessions that are neither admin, spider, or program. Web traffic continues to grow but SQL traffic stabilized at ~1,000 requests per day. The monthly traffic is smoothed by a 3 month moving window. Compare this figure to "all traffic" of Figure 3.

Table 4 shows the Web and SQL traffic from the most active institutions (with administrative traffic removed but including spiders and bots). Most traffic is from programs that spider the Website, or download data. The *unknown* address is caused by bugs in our logging software that did not record some IP addresses.

Table 5 shows the traffic counts when one parses each domain's organization name, looking for words like "university" or "college" or "school" or "district". It indicates that most traffic comes from colleges and universities.

| Table 4: Main user institutions and request volumes | | |
|---|---|---|
| Page Views | SQL | Institution |
| 4,668,124 | 3,114,078 | NASA |
| 3,933,370 | 104,378 | Google Inc. |
| 2,695,292 | 65,226 | Johns Hopkins University |
| 2,241,295 | 2,196,411 | AstroWise |
| 1,959,910 | 1,884,477 | NRC Canada |
| 1,943,511 | 816 | University of California |
| 1,261,638 | 971,166 | University of Illinois, CCSO |
| 1,168,071 | 70,628 | Microsoft Corp |
| 1,094,922 | 558 | Pino Torinese Observatory |
| 728,123 | 543,377 | Oxford University |
| 708,429 | 806,630 | Universidad de Cantabria |
| 644,986 | 458,636 | Max-Planck-Institut Garching |
| 455,061 | 390,805 | Inst. Astrofisica de Canarias |
| 14,969 | 770,019 | Unknown |

| Table 5: Traffic by Domain Name "type". | | | |
|---|---|---|---|
| type | institutions | page views | SQL |
| University | 863 | 31,507,386 | 8,648,855 |
| College | 407 | 478,996 | 1,410 |
| School | 310 | 823,138 | 1,890 |
| Other .edu | 169 | 7,554,956 | 3,509,361 |
| .gov | 238 | 446,460 | 83,562 |

## 4.2 Traffic by Most Popular "verbs"

Table 6 shows the importance of spatial data search for Astronomy applications. Of the 13.3M SQL queries, 5.8M involved spatial search functions (like fGetNearbyObjEq()), and all but one ("default") of the next 5 most popular verbs are variants of "get data near this point."

## 4.3 Traffic on Parts of Web Site

Again, subtracting out the admin and spider traffic, the traffic in the Website partitions approximately along the menu hierarchy of the site's home page (tools, get data, projects, help, ….) Table 7 gives the traffic breakdown by part of the Website. Most traffic goes to the tools that view the data and images. The third most popular part of the site is the astronomy educational activities with 4M page views and over 600k sessions.

**Table 6:** Most popular web "verbs".

| verb | page views | description |
| --- | --- | --- |
| x_sql | 13,393,187 | Ad hoc SQL query |
| default | 10,394,717 | Navigation page |
| GetJpeg | 8,929,524 | Get an object's image |
| x_radial | 6,023,717 | Radial DB search |
| x_rect | 5,673,636 | Rectangular search |
| showNearest | 3,388,016 | Nearest object to point |
| obj | 2,511,025 | Get Photo Object by ID |
| specById | 2,037,324 | Get Spectrogram by ID |
| OEtoc | 1,438,447 | Object Explorer root |
| camcol | 1,307,075 | Camera column (band) |
| shownavi | 1,169,273 | Visual Navigation Page |
| frameByRCFZ | 1,114,325 | Get Frame |

**Table 7**: Web site traffic by part of tree

| Page views | web tree |
| --- | --- |
| 43,486,090 | tools: to use the database |
| 5,482,295 | get: data and image retrieval |
| 4,242,788 | proj: Science education projects |
| 3,986,970 | help: documentation on data and site |
| 560,198 | sdss: about SDSS |
| 549,148 | astro: about astronomy |
| 68,995 | skyserver: about Sky Server |

## 4.4 Traffic by Language

Figure 7 shows the page view traffic aggregated by language (English, German, Hungarian, Japanese, Portuguese, and Spanish). The non-English traffic largely reflects people using the site to learn about the SDSS or using it for education. The recent dramatic rise in the German traffic after 4 years at 2k page views per month to 80k page views per month is due to a much better German translation of the project website. We are very pleased by the traffic growth in the Spanish, Portuguese, and Hungarian sites.

## 4.5 Traffic on the Educational Website

Of particular interest to us is the use of the Project Website that teaches astronomy. It received 4.2M page views in all. Table 8 shows that these page views are largely concentrated on the Advanced and Basic projects that teach astronomy. With bots and spiders excluded, there were 618K sessions involving at least one project page. The 297K sessions involving two or more project pages suggest that the student was reading the material rather than just browsing. Those sessions had 7.4 million page views, more than 21 thousand SQL queries, and delivered more than 47 thousand hours of instruction. Few astronomy textbooks or teachers can match that record.

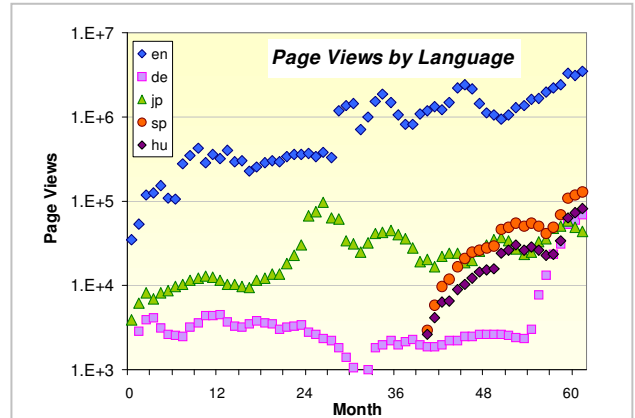

**Figure 7:** SkyServer Web page views (averaged over 3 month windows) for each language sub-site. Note the rapid growth in the Spanish, Hungarian, and Portuguese sites.

**Table 8:** Page views of Project website by area

| "area" | pg views | focus |
| --- | --- | --- |
| Advanced | 1,752,889 | Teaches astronomy. |
| Basic | 1,075,487 | Tells what astronomy is. |
| Kids | 489,438 | Very elementary. |
| Teachers | 364,553 | Advice to teachers. |
| Games | 224,888 | Hunt pictures for examples |
| Challenges | 112,019 | Some open ended projects |
| Links | 40,725 | Pointers to other places |
| Mailing | 11,006 | Talk to authors |
| High School | 3,733 | Grades 9-12 |
| Cool | 3,459 | Fun things. |
| User | 2,134 | User registration |
| Middle School | 840 | Grades 4-8 |
| Get Answers | 461 | Answers to exercises |
| Lower School | 410 | Grades K-4 |
| Evaluate | 225 | Comment on site. |

## 4.6 Traffic by Data Release

The SDSS has released six versions of the "Best" catalogs that contain all the data processed with the most recent software – these versions are called the Early Data Release and Data Releases 1 through 5 (EDR, DR1-5). In addition, there are Target DR1-5 containing the data snapshot used for selection of spectroscopic targets. There is also a Runs database that is mosaiced to produce the "Best" database. 99.98% of the traffic goes to the Best catalogs. DR5 was just coming online when the logs for this study were frozen. The SDSS Servers have answered about 16 million SQL queries when spider and admin queries are excluded.

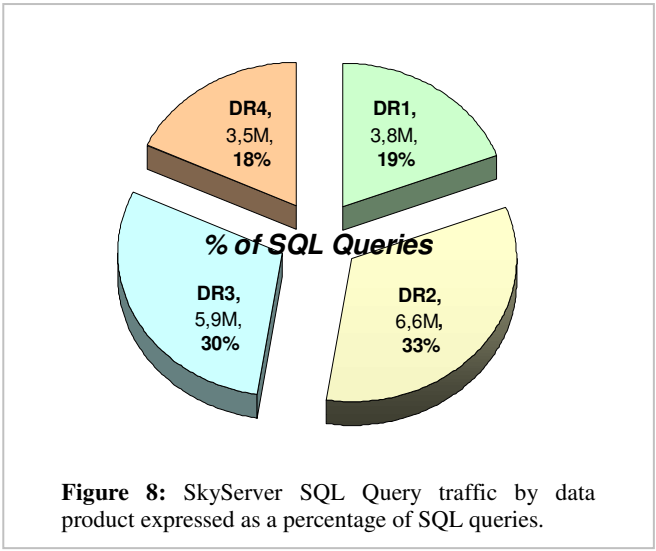

**Figure 8:** SkyServer SQL Query traffic by data product expressed as a percentage of SQL queries.

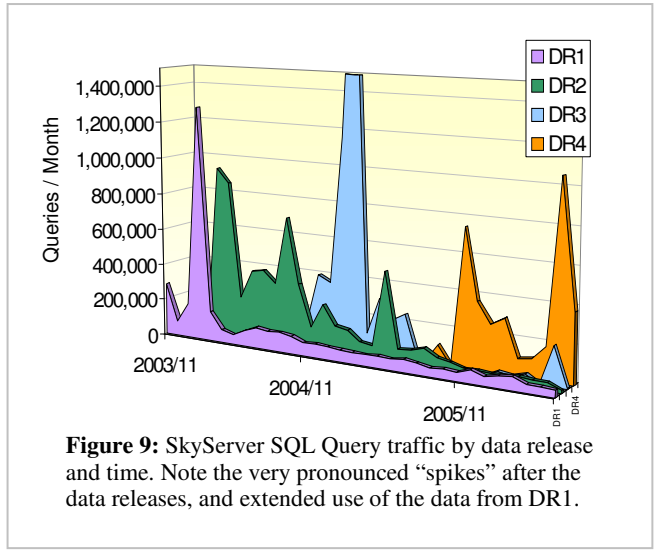

**Figure 9:** SkyServer SQL Query traffic by data release and time. Note the very pronounced "spikes" after the data releases, and extended use of the data from DR1.

Each catalog has been available for ad-hoc SQL queries. Figure 8 shows the relative traffic (measured in SQL queries) on the catalog. It shows that each dataset has had comparable traffic, although the newer ones have had less time to garner traffic – This reflects the increased interest in the datasets with time.

Figure 9 shows that in 2003 the SQL traffic quickly rose to about 10,000 queries per day and then held fairly steady at that saturation level. It also indicates that each product has received between 400K and 600K SQL queries. The newer products have had a shorter time to accumulate "hits." The figure also shows that interest in a product "spikes" soon after it is released and then declines as newer products are released. The huge spike in DR3 is a substantial copy of the database by one site. The oldest product, DR1, still gets about 40K SQL queries per month – there are still people doing science with it. This shows that once a data product is published, it needs to stay online forever – much as scientific literature must remain available to allow others to read, verify and extend previous work.

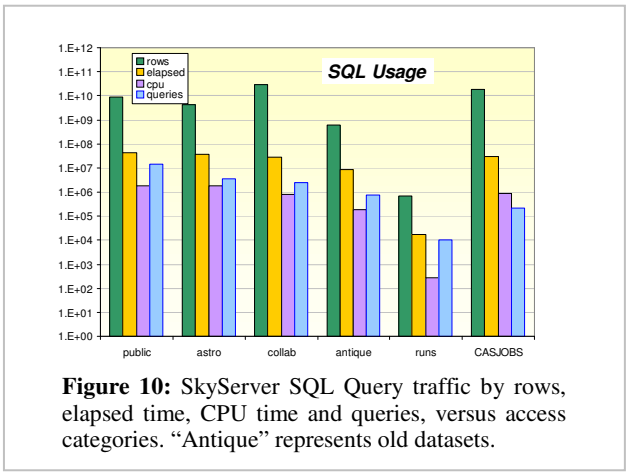

**Figure 10:** SkyServer SQL Query traffic by rows, elapsed time, CPU time and queries, versus access categories. "Antique" represents old datasets.

### 4.7 Traffic by Professional Astronomers

The SkyServer Website is the easiest way to access the Sloan Digital Sky Survey catalogs. There are three services set up for professional Astronomers. Some "collaboration" servers are set up for the exclusive use of members of the SDSS collaboration that allow early access to the data for peer review, allow them to run larger queries with larger answer sets, and provide a more spartan user interface. There is also a public service "/astro/" with the same /collab/ spartan interface. In addition, there is a public CasJobs (Catalog Archive Service Batch Jobs) interface that allows users to create a personal database on the server (MyDB), upload data to it, use data from that database in queries, and send results of queries to that database. Section 7 discusses CasJobs.

The /collab site delivered 3.4M page views and 2.4M SQL queries, the /astro site delivered 9.2M page views and 3.6M SQL queries, and /CasJobs 1.1 delivered 1.1M page views and executed 209K jobs. Put another way, the public sites got three times more traffic than the collaboration site, indicating that the datasets are widely used and that people who were not "insiders" were able to use the data (see Figure 10.) We will return to the kinds of SQL queries the professionals presented in sections 6 and 7.

### 4.8 Traffic by Server

As of mid-2006, there were 36 Web server logs and 15 SQL server logs being harvested. Three servers handle 74% of all Web requests and the top 3 SQL servers service 65% of all SQL requests.

The public SkyServer has a 10-minute 100K row limit, the astro and collab sites have a 1-hour 500K row limit, and the CasJobs long queue has an 8 hour. The CasJobs collab site has a 24 hour.

The Web servers are lightly loaded except for the compute-intensive task of composing JPEG images on demand from the images in the database (the GetJpeg verb of Table 5). That requires retrieving and converting several Jpegs from the database into a bitmap, decorating the bitmap with information fetched from the database, clipping the result to an image centered on the desired spot and then converting the bitmap to a JPEG image and delivering it via SOAP over HTTP. This entire image-processing task can consume 0.3 to 5.0 CPU seconds depending on the image size and query complexity.

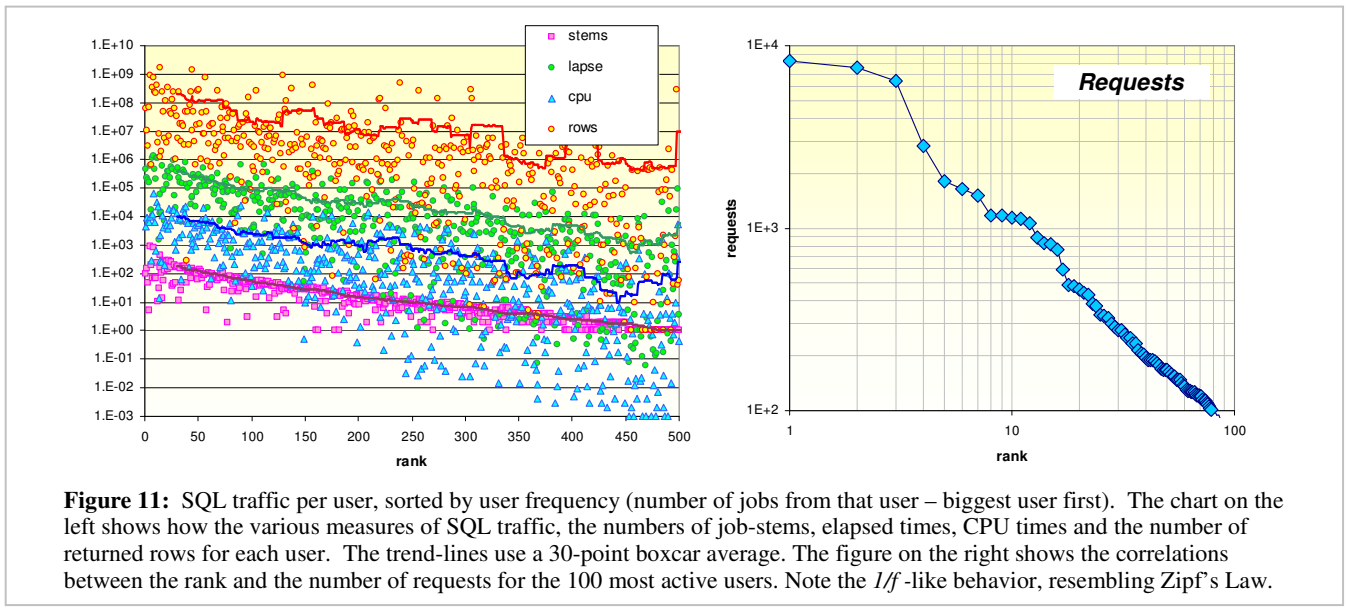

**Figure 11:** SQL traffic per user, sorted by user frequency (number of jobs from that user – biggest user first). The chart on the left shows how the various measures of SQL traffic, the numbers of job-stems, elapsed times, CPU times and the number of returned rows for each user. The trend-lines use a 30-point boxcar average. The figure on the right shows the correlations between the rank and the number of requests for the 100 most active users. Note the *1/f* -like behavior, resembling Zipf's Law.

Fermilab has a load-balanced pool of three Web servers allocated to service these tasks and the less demanding ones on a round-robin basis (Figure 1). The average network traffic on these servers is 1.5 MB/s (~10 Mbps) – but peak traffic can be whatever is available. Outbound bandwidth is ~100x inbound bandwidth.

# 5 CasJobs and MyDB – Batch is Back

CasJobs (Catalog Archive Server Batch Jobs) is a Web interface and a Web-service SOAP interface that allows users to submit unrestricted query scripts in batch mode to the SDSS databases, create a personal database on the server (MyDB), upload data to it, use data from that database in queries, and send results of queries to that database. CasJobs is described in detail in [3, 5] and is on the Web for you to explore. Login is required to submit a background job or create a database, but anyone may create an account.

As described in [5], the primary motivations for CasJobs were to

- separate the quick queries (that finished in under a minute) from the long queries that took minutes or hours to execute and bogged the server down for everyone else;
- provide each user with a server-local scratch workbench database to hold intermediate query results and *bring the analysis to the data* as much as possible rather than vice-versa (this avoids moving large intermediate data results across the Internet); and
- provide simple load balancing among the servers by distributing the queues across the server pool.

## 5.1 CasJobs Queries

CasJobs processed 209K jobs since September 2004. The CasJobs Administrator did 60% of these jobs (scheduling jobs and reporting on them). In what follows, we exclude the administrator. This leaves 77K jobs to consider, of which 59K are syntactically correct.

There have been 537 distinct users. Virtually all "big" queries were IO bound (disk bound). CPU utilization rarely got above 10% of the elapsed time. Faster CPUs would not help, but more RAM and more disks would have sped the queries. Put another way, the total CasJobs elapsed time was slightly more than one year and the total CPU time was less than 10 days (229 hours).

Figure 11 displays the CasJobs activity (jobs, elapsed time, CPU time, rows returned) of all 537 users sorted by the number of jobs they submitted. It shows no clear breakpoints: some users submitted a few big jobs, some users submitted many big jobs, some submitted many small jobs, many submitted a few big jobs, and many submitted a few small ones. When sorted by jobs, the top 10 users submitted 57% of the jobs; sorted by time, the top 10 used 27% of the time, and sorted by rows returned, the top 10 retrieved 51% of the rows. When we rank users by the number of requests they submitted, the rank vs the requests shows a nice power-law behavior, resembling a *1/f* distribution, indicating that users do not have a characteristic workload scale.

Figure 11 also shows that CPU time (blue) is much less than wait time (green) in almost all cases – even though these are 2-way or 4-way multiprocessor servers. It shows that the system returns about 20K rows per CPU second and 500 rows per elapsed second. CasJobs returned 16 billion rows in all (about 1TB). This is twice the number of rows returned by the public server, four times the number returned by the /astro server, but only 60% of the number returned by the /collab servers (see also Figure 10).

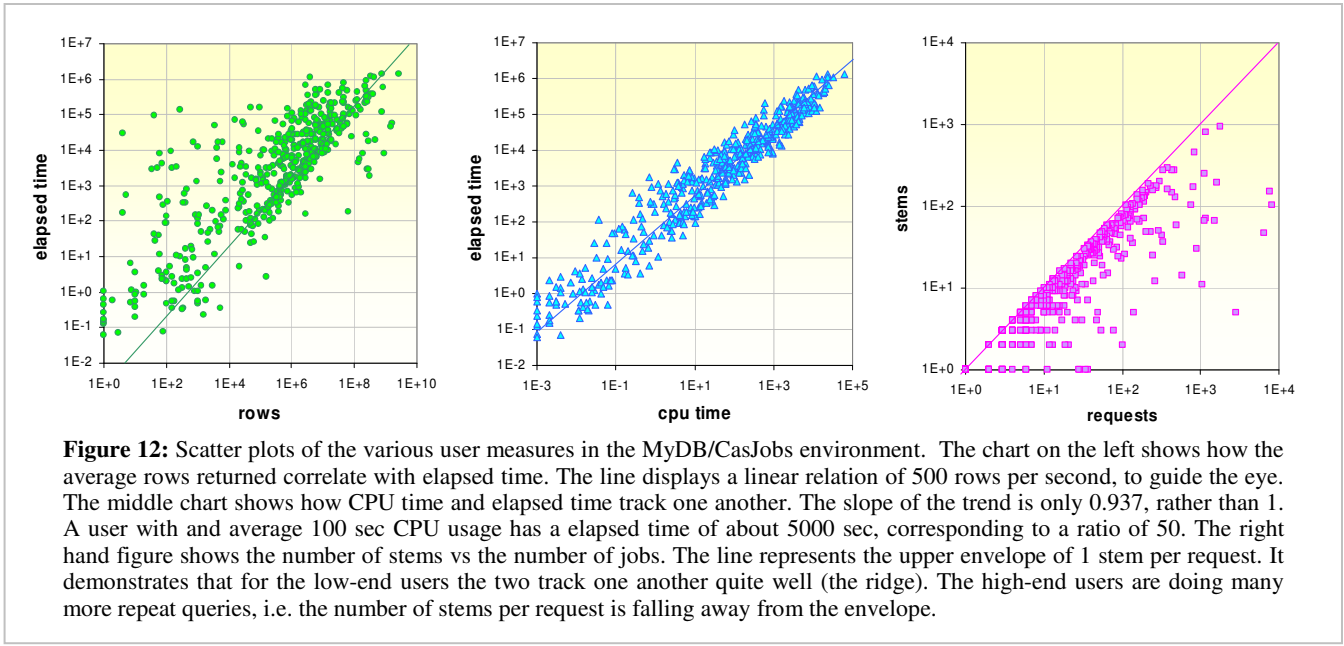

**Figure 12:** Scatter plots of the various user measures in the MyDB/CasJobs environment. The chart on the left shows how the average rows returned correlate with elapsed time. The line displays a linear relation of 500 rows per second, to guide the eye. The middle chart shows how CPU time and elapsed time track one another. The slope of the trend is only 0.937, rather than 1. A user with and average 100 sec CPU usage has a elapsed time of about 5000 sec, corresponding to a ratio of 50. The right hand figure shows the number of stems vs the number of jobs. The line represents the upper envelope of 1 stem per request. It demonstrates that for the low-end users the two track one another quite well (the ridge). The high-end users are doing many more repeat queries, i.e. the number of stems per request is falling away from the envelope.

CasJobs saved bandwidth and user wait time. When it was introduced, the public "big queries" moved to CasJobs. CasJobs allows users to do sophisticated and high-bandwidth analysis near the server without having to provision their own server and without having to download the SDSS archive to their site. Of the 59K valid CasJobs, 85% are simple selects, about 9K are complex programs that create tables in MyDB, define local variables, constants, and functions, and then do multi-step analysis of the SDDS catalogs and the user's MyDB data.

## 5.2 CasJobs Users

Figure 12 shows scatter plots between the various per-user average measures and approximate trend lines to emphasize characteristic correlations. The left hand figure shows the number of rows returned vs elapsed time. The trend line corresponds to a 750 rows per sec average. It is obvious, that the scatter is quite large. For an elapsed time of about 100 sec one can see the sharp cutoff at around 500,000 rows, also noticeable on the right hand plot on Figure 11.

Figure 12 shows that CPU time and elapsed time track one another quite well, but the slope is less than one. The characteristic ratio is about 34 sec of elapsed time for 1 sec of CPU, as measured on the high end, compared to a factor of 100 for short queries. This indicates that most of the workload is heavily I/O bound.

High end users tend to use a smaller number of "stems" than the low-end users. This trend is clearly seen from the third panel on Figure 12: with more requests, the number of stems is falling away from the envelope. The meaning of this is that the people with lots of jobs are probably refining a complex query, or doing a spatial search by changing some the parameter values, and running a similar query pattern many times. This vindicates one of our main arguments for CasJobs – to avoid wasting bandwidth from repeat queries during the refinement process.

The average query is about 100 tokens long, runs for about 10 minutes, uses 15 CPU seconds and deposits 250K rows in MyDB. Since CasJobs was inaguratead three years ago, traffic has been fairly steady at about 10 queries per hour. There have been peak periods of 700 queries per hour; but, on average jobs are processed within a few hours. SQL

## 6 Query Analysis

Of the 20.7M SQL queries, there were 10.3M distinct queries, 9.0M of these statements are syntactically correct, and 7.4M returned at least one row suggesting they were valid queries. When one replaces all the numbers in a query with a "#" symbol, the set of queries shrinks to 138K templates of which 102K are syntactically correct and 78K return results.

### 6.1 Queries from Bots

Robot sessions show up with very few templates compared to the number of SQL queries – typically thousands of queries in a session with just one template. If we say that a session with the same template repeated 4 or more times (`session.SqlStem>4*session.sql`) represents a robot or program, then those 10.9K robot sessions represent 15.7M of the SQL queries with only 12K SQL templates – the typical bot is reissuing the template 13K times! The residue 85.8K sessions submitted 417K SQL queries.

| Table 9: Most popular function calls in SQL queries. | |
|---|---:|
| verb | queries |
| fGetNearbyObjEq | 8,698,330 |
| fGetObjFromRect | 3,269,000 |
| fGetNearestObjEq | 661,349 |
| fGetUrlFitsCFrame | 88,453 |
| fGetNearestFrameEq | 78,625 |
| fGetNearestObjIdAllEq | 56,063 |
| fGetNearestObjIdEqType | 18,052 |
| fGetUrlFitsField | 9,016 |
| fGetObjFromRectEq | 5,638 |

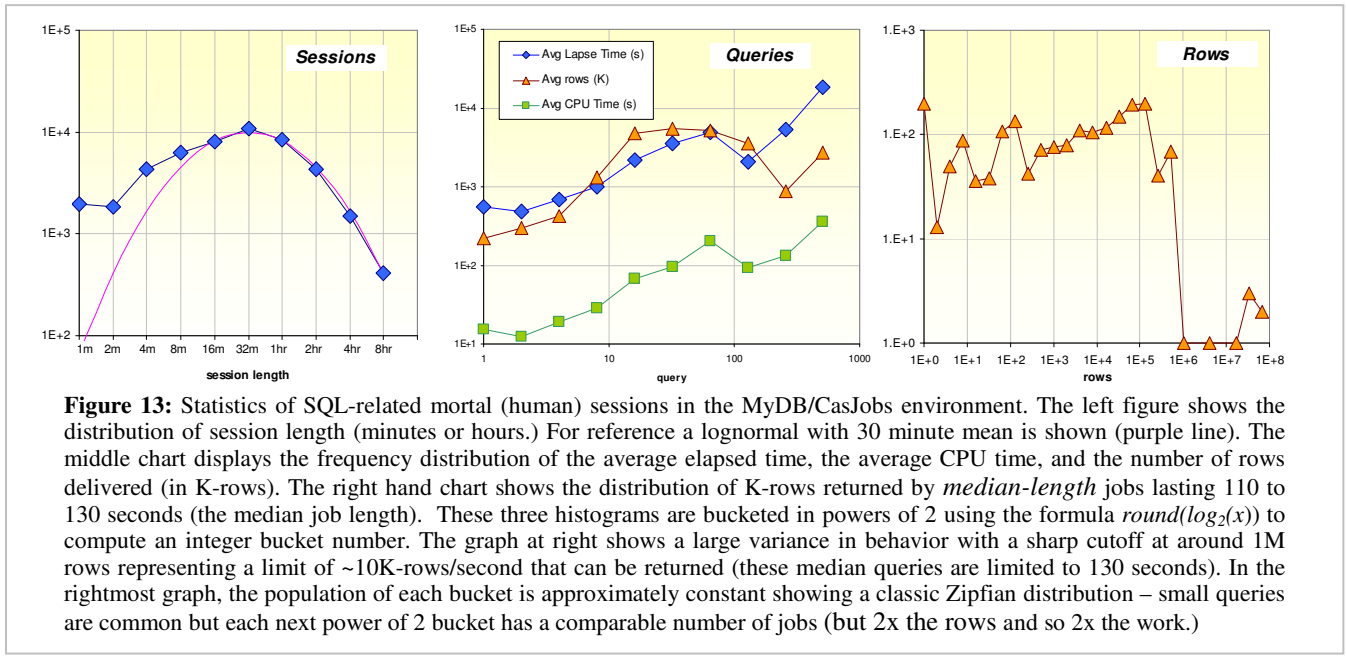

**Figure 13:** Statistics of SQL-related mortal (human) sessions in the MyDB/CasJobs environment. The left figure shows the distribution of session length (minutes or hours.) For reference a lognormal with 30 minute mean is shown (purple line). The middle chart displays the frequency distribution of the average elapsed time, the average CPU time, and the number of rows delivered (in K-rows). The right hand chart shows the distribution of K-rows returned by *median-length* jobs lasting 110 to 130 seconds (the median job length). These three histograms are bucketed in powers of 2 using the formula $round(log_2(x))$ to compute an integer bucket number. The graph at right shows a large variance in behavior with a sharp cutoff at around 1M rows representing a limit of ~10K-rows/second that can be returned (these median queries are limited to 130 seconds). In the rightmost graph, the population of each bucket is approximately constant showing a classic Zipfian distribution – small queries are common but each next power of 2 bucket has a comparable number of jobs (but 2x the rows and so 2x the work.)

The robots typically do a spatial search. Table 9 shows the counts for the most popular functions. All but 2 of the functions in Table 9 are spatial data lookups. Many other robot queries systematically vary the parameters of an `RA,DEC` bounding box using the SQL `BETWEEN` construct. 610 of the bot templates (~5%) have that construct. 10K of the residue have that construct -- about 12% of the 74K valid templates from sessions that seem not to be bots. After failing to "teach" users to use the spatial search functions, we added an RA-dec index to speed this bounding-box construct.

### 6.2 Queries from Mortals

Let us try to characterize non-bot SQL queries. Define *mortal queries* as ones that are in a session where the number of distinct SQL templates is at least 20% of the number of SQL queries (that is, the typical query is not re-used more than 4 times) and where the session is less than 8 hours. Further define *valid mortal queries* as those that return at least one row from the database. Let's analyze these mortal queries and their sessions.

There are 85k mortal sessions with 412K queries, of which 271K (66%) are valid mortal queries. The typical session has six SQL queries and lasts thirty minutes – but sessions of four hours are quite common (see Figure 11). The median valid query ran for two minutes (127 seconds) and those queries had a median of 2 seconds of CPU time and 3.5K rows returned. As Figures 11, 12 and 13 indicate, these numbers have *huge* variance – the median and average are very different. The average number of rows returned was 187K not 3.5K.

The 271K valid mortal queries also have a wide range of complexity. There are 74K valid mortal templates. Of those, 71% use the *select-from-where* syntax, 14% use the *select-from-where-orderBy* syntax, and 6% are *select* from a table-valued function – so 91% follow that simple *select-from* format. But some queries have more than 80 *select* clauses; some have 7 *group-by* clauses and there is considerable use of *outer-joins* and many other advanced features. Approximately 13k templates (18%) involve a spatial join using one of the table valued spatial functions. The numbers are probably somewhat skewed by a set of more than 50 sample queries that are available on the SkyServer Help page [15]. Many users tend to initially run these either with or without modification until they get proficient enough to formulate their own queries.

### 6.3 Term Frequency within SQL Queries

As in Abdulla [12], we analyzed token frequencies within the SQL templates. We simplified the SQL query templates by removing parentheses, table aliases, database and table prefixes, and function parameter names. We also substituted tokens for strings, comparison operators (e.g. '>=', *between*), bitwise operators, arithmetic operators, logical operators, and for multi-word SQL keywords such as *group by* and *order by*. This produces about 110K query templates. The templates mention all 44 tables, but 493 of the 2,228 columns are never mentioned and 36 of the 109 built in functions are not used.

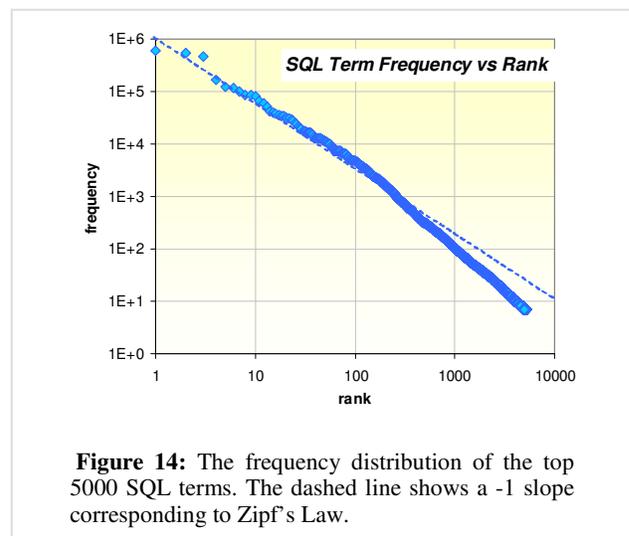

**Figure 14:** The frequency distribution of the top 5000 SQL terms. The dashed line shows a -1 slope corresponding to Zipf's Law.

SQL is a formal language so one might not expect to see a Zipfian distribution of term frequency so characteristic of natural languages. But, indeed that is what we see. Ignoring term context (part of speech) and spelling errors, and plotting term rank vs frequency gives Figure 14 which indeed looks like a simple power law.

Looking deeper into the language, separating SQL verbs, column names, and table names gives the top-30 frequency counts. Table 10 shows the top 30 SQL token frequencies, Table 11 shows the top 30 table frequencies and Table 12 shows the frequency of the top 30 columns. The full data is graphed in Figure 15. The Figure and the Tables indicate that the distributions are not exactly Zipfian, but term frequencies do decline very sharply. The "staircase" effect is caused by correlated constructs like *select-from-where* and *case-when-then-end*.

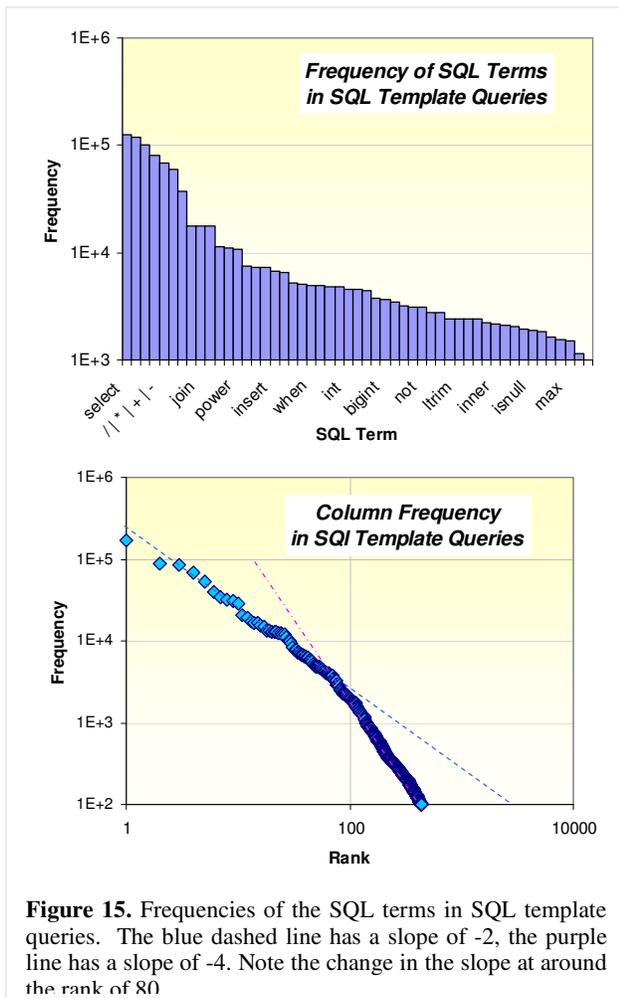

**Figure 15.** Frequencies of the SQL terms in SQL template queries. The blue dashed line has a slope of -2, the purple line has a slope of -4. Note the change in the slope at around the rank of 80.

## 6.4 Using Templates to Correct SQL Queries

Query repetition may offer a way to improve the user experience. If a few of the 100K query templates are similar to a new user query has a syntax error, it might be useful to offer a similar correct queries from the corpus. A simple distance function measures query similarity: First chop each template into token substrings (N-grams) [16] that are then sorted. Then compute the Jaccard distance [17] betweeen the query's N-gram sets and the N-grams of each template. This finds near matches to a user's query. The templates and actual correct query examples can be returned as suggestions. For example, consider the following incorrect SQL query:

```
SELECT TOP 10 ph.ra,ph.dec,
       str(ph.g - ph.r,11 ? ) as color,
       isnull(s.bestObjId, 0) as bestObjId,
       'ugri'
FROM #x x, #upload up,
     BESTDR2..PhotoObjAll as ph
LEFT OUTER JOIN ? SpecObjAll s
  ON ph.objID = s.bestObjID
WHERE (ph.type=3 OR ?)
AND up.up_id = x.up_id
? x.objID=p
?.objID
ORDER BY x.up_id
```

The red question marks denote syntax errors. We passed this query into the matching system and got back the top three matches in Table 10.

The top query result fills in the missing values for our input query exactly. Also notice how the next two candidates

| Table 10: Three correct matching queries. ||
|---|---|
| SQL Query | Similarity |
| `SELECT TOP 50 p.ra,p.dec,`<br>`  str(p.g - p.r,11,8) as grModelColor,`<br>`  isnull(s.bestObjID,0) as bestObjID,`<br>`  'ugri' as filter`<br>`FROM  #x x, #upload u,`<br>`     BESTDR2..PhotoObjAll as p`<br>`LEFT OUTER JOIN BESTDR2..SpecObjAll s`<br>`  ON p.objID = s.bestObjID`<br>`WHERE ( p.type = 3 OR p.type = 6)`<br>`  AND u.up_id = x.up_id`<br>`  AND x.objID=p.objID`<br>`ORDER BY x.up_id` | 74% |
| `SELECT TOP 50 p.ra,p.dec,`<br>`p.run,p.rerun,p.camCol,p.field,p.obj,`<br>`  isnull(s.ra,0) as ra,`<br>`  isnull(s.[dec],0) as [dec],`<br>`  'ugriz' as filter`<br>`FROM  #x x, #upload u,`<br>`     BESTDR2..PhotoObjAll as p`<br>`LEFT OUTER JOIN BESTDR2..SpecObjAll s`<br>`  ON p.objID = s.bestObjID`<br>`WHERE ( p.type = 3 OR p.type = 6)`<br>`  AND u.up_id = x.up_id`<br>`  AND x.objID=p.objID`<br>`ORDER BY x.up_id` | 66% |
| `SELECT TOP 50 p.ra,p.dec,`<br>`p.run,p.rerun,p.camCol,p.field ,p.obj,`<br>`  isnull(s.ra,0) as ra,`<br>`  isnull(s.[dec],0) as [dec],`<br>`  'ugriz' as filter`<br>`FROM  #x x, #upload u,`<br>`     BESTDR2..PhotoObjAll as p`<br>`LEFT OUTER JOIN BESTDR2..SpecObjAll s`<br>`  ON p.objID = s.bestObjID`<br>`WHERE  ( p.type = 3 OR p.type = 6)`<br>`  AND u.up_id = x.up_id`<br>`  AND x.objID=p.objID`<br>`ORDER BY x.up_id` | 60% |

follow the same TOP N, temporary table, LEFT OUTER JOIN sequence, and WHERE conditional syntax usage. Since we record error messages for each SQL query we only present correct queries. This example illustrates template similarity and the large corpus of templates can provide suggestions to users.

### 6.5 Examples of Complex SQL Queries

About 8K templates have explicit join verbs. Multi-way complex joins are common. The following 8-way join is typical:

```sql
SELECT LF.BESTOBJID, LF.TARGETID
FROM MYTABLE_61          AS LF
INNER JOIN PHOTOTAG      AS BP
       ON LF.BESTOBJID = BP.OBJID
INNER JOIN TARGETINFO    AS TI
       ON TI.TARGETID = LF.TARGETID
INNER JOIN PHOTOTAG      AS TP
       ON TI.TARGETOBJID = TP.OBJID
INNER JOIN FIELD         AS TF
       ON TF.FIELDID = TP.FIELDID
INNER JOIN SEGMENT       AS TS
       ON TS.SEGMENTID = TF.SEGMENTID
INNER JOIN FIELD         AS BF
       ON BF.FIELDID = BP.FIELDID
INNER JOIN SEGMENT       AS BS
       ON BS.SEGMENTID = BF.SEGMENTID
LEFT OUTER JOIN SPECOBJ AS SO
       ON LF.BESTOBJID = SO.BESTOBJID
```

Another interesting example is this 16-way join:

```sql
select count_big(distinct g.objid)
from PhotoObjAll as g
left outer join PhotoProfile as p0
on g.objId=p0.objID
left outer join PhotoProfile as p1
on g.objId=p1.objID
left outer join PhotoProfile as p2
  repeated to 15 times for each p(i)…
left outer join PhotoProfile as p14
on g.objId=p14.objID
where g.run = # and g.rerun = #
  and g.camcol = # and g.field = #
  and g.obj != #
  and ((p0.bin=# and p0.band=#)or(p0.bin is null))
```
*repeated 15 times for each p(i)*

There is an 85-way *union*! There are complex *sub-selects* nested 7 deep. In general, some of the users are very ingenious, and some have SQL skills that qualify them as database gurus.

### 7 Summary

These results are tantalizing. Each answer suggests other questions. A few key patterns emerge from this forest of data. SkyServer traffic nearly doubled each year – both Web traffic and SQL queries grew by about 100%/year. We failed to find clear ways to segment user populations. We were able to ignore the traffic that was administrative or was eye-candy, leaving us with a set of 65M page views and 16M SQL queries. We organized these requests into about 3M sessions, about half of which were from spiders. The residue of 1.5M sessions had 51M page views and 16M SQL queries – still a very substantial corpus.

Our best estimate is that spiders contributed 46% of sessions and 20% of the Web traffic. Scientific analysis programs and data downloaders were 3% of the sessions, but 37% of the Web traffic and 88% of the SQL traffic. Interactive human users were 51% of the sessions, 41% of the Web traffic and 10% of the SQL traffic. We cannot be sure of those numbers because we did not find a very reliable way of classifying bots vs mortals.

The human traffic seems to grow a little slower than the whole. The yearly growth is still exponential, but the traffic only doubles every 1.33 years.

Many of our logs exhibit a remarkable power law behavior. It is well-known that long-tailed distributions emerge naturally from multiplicative processes [18, 25, 26], when the product of many factors determines the final outcome. It has been pointed out recently [19, 20, 28] that such behavior is also natural in social networking, especially so in Web-based systems where users are presented with many choices. We find such long-tailed distributions in the page views and the lengths of sessions, and also in the number of SQL requests. Some of these power laws extend the *1/f* behavior over 6 orders of magnitude (e.g., Figure 5a).

One thing that is clear is that there is considerable interest in the educational site in each of the five available languages. There were 297K sessions involving two or more project pages with behavior that "looked" mortal. Those sessions had 7.4 million page views, more than 21 thousand SQL queries, and delivered more than 47 thousand hours of instruction. Few astronomy textbooks or teachers can match that record.

The SkyServer will write some SQL for you – and many users used the fill-in-the-form user interface – but hundreds of astronomers "graduated" to the free-form SQL query interface where they composed tens of thousands of SQL queries, and about 500 astronomers have created their own private database and run complex analysis jobs using the CasJobs site. There was considerable skepticism whether this would work at all, whether it would be useful, and whether it would be abused. So far it has been quite useful to some and has not been abused. The CasJobs template in fact has been successfully adopted by other astronomical archives like GALEX [21], and even non-astronomical archives like AmeriFlux [22].

In terms of interest in the data, each new data release gets a flurry of interest. First there is early mortal traffic, then there is an intense period of bot (program) download and analysis, and after that (when a new version appears) traffic subsides to a few thousand queries per month. So far no release has gone out of use. This confirms our belief that once published, scientific data must remain online and accessible so that scientists can repeat experiments or analyses indefinitely. The fact that earlier releases like DR1 continue to get sustained usage is of especial significance for the budgeting of data access resources for the next generation of large astronomical surveys like Pan-STARRS [23] and LSST [24].

SkyServer is an example of the new way to publish and access scientific data. It is the data and documentation produced by a collaboration along with tools to analyze the data. It is public, and it can be federated with other scientific archives and with the literature. We hope that it will turn into a useful resource for more complex analyses by others than those presented in this paper.

## 8 Acknowledgments

This research was enabled by the dataset created by the Sloan Digital Sky Survey and by the website that hosts the data. So, this work owes a huge debt to the astronomers who designed and built the telescope, who gathered the data, who wrote the software to convert pixels into the SDSS catalog, and who ran those pipelines. Many others built the SkyServer website, CasJobs, and other web services. Others developed educational materials using the data and have translated the website into 5 different languages. This article would not have been possible without all those contributions.

Funding for the SDSS and SDSS-II has been provided by the Alfred P. Sloan Foundation, the Participating Institutions, the National Science Foundation, the U.S. Department of Energy, the National Aeronautics and Space Administration, the Japanese Monbukagakusho, the Max Planck Society, and the Higher Education Funding Council for England. The SDSS Web Site is http://www.sdss.org/.

The SDSS is managed by the Astrophysical Research Consortium for the Participating Institutions. The Participating Institutions are the American Museum of Natural History, Astrophysical Institute Potsdam, University of Basel, University of Cambridge, Case Western Reserve University, University of Chicago, Drexel University, Fermilab, the Institute for Advanced Study, the Japan Participation Group, the Johns Hopkins University, the Joint Institute for Nuclear Astrophysics, the Kavli Institute for Particle Astrophysics and Cosmology, the Korean Scientist Group, the Chinese Academy of Sciences (LAMOST), Los Alamos National Laboratory, the Max-Planck-Institute for Astronomy (MPIA), the Max-Planck-Institute for Astrophysics (MPA), New Mexico State University, Ohio State University, University of Pittsburgh, University of Portsmouth, Princeton University, the United States Naval Observatory, and the University of Washington.

Alex Szalay acknowledges support from NSF AST-0407308, from the Gordon and Betty Moore Foundation and the W.M. Keck Foundation. We benefited from discussions with Tanu Malik and Stuart Ozer about SQL query templates and their statistics. Conversations with Raul Singh and Ghaleb Abdulla, and Richard Lees about their SkyServer log analysis were also very useful. Mark Manasse was very helpful in discussions that led to the template-matching ideas in section 6.5.